\documentclass[doublecol]{epl2} 

\usepackage{amsmath}

  \title{The Kuramoto model with distributed shear}

\author{Diego Paz\'o\inst{1} \and Ernest Montbri\'o\inst{2}}
 \shortauthor{Diego Paz\'o and Ernest Montbri\'o}

\institute{                    
  \inst{1} Instituto de F\'{i}sica de Cantabria (IFCA), CSIC-Universidad de
Cantabria - 39005 Santander, Spain \\
  \inst{2} Department of Information and Communication Technologies,
Universitat Pompeu Fabra - 08003 Barcelona, Spain
}

\abstract{We uncover a solvable generalization 
of the Kuramoto model in which shears (or nonisochronicities) 
and natural frequencies are distributed and statistically dependent.
We show that the strength and sign of this dependence greatly alter 
synchronization and yield qualitatively different phase diagrams.  
The Ott-Antonsen ansatz allows us to obtain analytical
results for a specific family of joint distributions.
We also derive, using linear stability analysis,
general formulae for the stability border of incoherence.
}

\begin{document}

\maketitle

\section{Introduction}
Collective synchronization is a commonly observed phenomenon in nature and
in some technological applications \cite{Win67,Kur84,PRK01,MMZ04,ABP+05},
in which mutual interactions succeed to
entrain the rhythms of a heterogeneous ensemble of self-sustained oscillators.
It is mathematically captured by a prototypic minimal model
put forward by Kuramoto more than thirty years ago \cite{Kur75,Kur84}.
Moreover, this model is a suitable framework for the
quantitative analysis of a variety of physical systems such as arrays of 
Josephson junctions \cite{pere} or mechanical rotors/oscillators \cite{uchida,mertens,cross}.

The universal form of a limit-cycle close to a Hopf bifurcation
led Kuramoto to analyse collective synchronization 
resorting to the 
mean-field version of the complex Ginzburg-Landau equation
with disorder
\cite{Kur75,aizawa76,cross}:
\begin{equation}
\dot{z}_j = z_j [1 + {\rm i}(\omega_j + q_j) - ( 1 + {\rm i} q_j) |z_j|^2]+\frac{K}{N}
\sum_{k=1}^{N} (z_k- z_j),
\label{Complex2}
\end{equation}
where $z_j=\varrho_j {\rm e}^{{\rm i}\theta_j}$, and $j=1,\ldots,N\gg1$. Here,
$\omega_j$ is the natural frequency of the $j$-th oscillator, whereas 
$q_j$ is the so-called shear (or nonisochronicity) that quantifies the dependence of
the oscillation frequency on the amplitude.
Under the assumptions that the coupling is purely diffusive ($K$ real) and 
weak ($|K|$ small), a phase reduction of eq.~(\ref{Complex2})  yields \cite{Kur84}
\begin{eqnarray}\label{model}
  \dot{\theta_j}  = \omega_j + K  q_j + \frac{K}{N} \sum_{k=1}^{N}
  \left[\sin(\theta_k-\theta_j)-q_j\cos(\theta_k-\theta_j)\right] .
\end{eqnarray}
We may also cast eq.~(\ref{model}) in a more compact form:
\begin{equation}
 \dot{\theta_j}  =  \omega_j + K  \tan \beta_j  
- \frac{1}{\cos \beta_j}\frac{K}{N} \sum_{k=1}^{N}
\sin(\theta_j-\theta_k+\beta_j) 
\label{model2}
\end{equation}
with  $\tan \beta_j=q_j$ and $|\beta_j|\le\frac{\pi}{2}$. 
Under the simplifying assumption that the
shears are not distributed, $q_j=\hat q$ ($\Rightarrow\beta_j=\hat \beta$),
the so-called Sakaguchi-Kuramoto model \cite{SK86} is recovered
(redefining $\omega_j'=\omega_j+K\tan\hat\beta$, and $K'=K/\cos\hat\beta$).
Additionally, 
under the more severe constraint $q_j=0$, 
eq.~(\ref{model2}) becomes the standard Kuramoto model \cite{Kur75}.

The goal of this work is to perform a detailed analysis of 
phase equations~(\ref{model}) under the assumption that
the natural frequency and the shear of each oscillator are drawn from
a joint probability density function (PDF), $p(\omega,q)$.
A particular case of this problem has been recently analysed
assuming the parameters $\omega$ and $q$ to be 
independent random variables, $p(\omega,q)=g(\omega)h(q)$  \cite{MP11}. 
An interesting finding is that, 
if the width of the distribution 
$h(q)$ exceeds a precise threshold,
diffusive coupling is unable to
counteract shear heterogeneity leading to a self-organized, 
synchronous state.  
This result is in sharp contrast with the well-known prediction 
of the Sakaguchi-Kuramoto ---or the Kuramoto--- model, where
collective synchronization is assured at 
large enough $K$ values.

How do these results translate into the case  
where natural frequencies and shears are statistically dependent?
This is the case one may encounter when studying the  
synchronization of any particular class of self-sustained oscillators. 
Generally, model-specific parameters affect  
\emph{both} the oscillator's natural frequency and shear. 
Therefore, heterogeneity in certain parameters will also result into 
heterogeneity of $\omega$ and $q$ with some 
functional or statistical dependence between
them; see {\it e.g.}, eqs.~(7) and (8) in \cite{pere} for such a situation,
though the heterogeneity of $q$ is eventually neglected to simplify the analysis.  
In previous work,
parameter dependencies were found to 
influence the effect of diffusive coupling on the variance of 
the ensemble's oscillator frequencies, a phenomenon 
called `anomalous phase synchronization' \cite{BMK03}. 

In this Letter, we 
define a conditional probability
of $\omega$ given $q$, $g_c(\omega|q)$, such that
$p(\omega,q)= h(q)g_c(\omega|q)$.
The marginal PDF $h(q)$ is assumed to be unimodal, symmetric and centred at $q_0$.
Additionally,
the conditional probability $g_c$ is chosen to be unimodal and of the form
$g_c(\omega|q)=g_c(\omega-mq)$. This restricts our results to a particular class
of distributions that is nonetheless wide enough to 
illustrate a number of different synchronization scenarios 
(particularly depending on the sign of $m$).

\section{Continuum limit}
In our theoretical analysis 
we neglect finite-size effects and consider the thermodynamic limit
$N\rightarrow \infty$ of model (\ref{model}).
It is possible then to drop the indices and define
the probability density for the phases $f(\theta,\omega,q,t)$. 
Thus $f(\theta,\omega,q,t)\, {\rm d}\theta \, {\rm d}\omega \, {\rm d}q$
is the ratio of oscillators at time $t$ with phases between 
$\theta$ and $\theta+{\rm d}\theta$,
natural frequencies between $\omega$ and $\omega+{\rm d}\omega$, and shear 
between $q$ and $q+{\rm d}q$.
The density function $f$ obeys the continuity equation 
 \begin{equation} \label{continuity}
\partial_t f = - \partial_\theta
\left( \left\{\omega+Kq+  \frac{ K}{2{\rm i}}
\left[r {\rm e}^{-{\rm i} \theta}(1 - {\rm i} q) - {\rm c.c.}\right] \right\} f \right),
 \end{equation}
where c.c.~stands for complex conjugate of the preceding term, 
and the complex order parameter $r$ is
\begin{equation}
r(t)\equiv R {\rm e}^{{\rm i}\Psi}= \iint_{-\infty}^{\infty} \int_0^{2\pi} {\rm e}^{{\rm i} \theta}
f(\theta,\omega,q,t)~ {\rm d}\theta ~ {\rm d}\omega ~ {\rm d}q.
\label{z}
\end{equation}
The mean field $r$ measures the degree of synchronization of the system. If the
oscillators are uniformly distributed, a state commonly referred to as incoherence,
$f(\theta,\omega,q,t)$ equals $p(\omega,q)(2\pi)^{-1}$, and  $r$ vanishes.
 States for which part or all of the population is entrained
at a given frequency result in a nonuniform distribution of the
phases such that $R>0$.

The density function $f(\theta,\omega,q,t)$ is real and $2\pi$-periodic 
function in the $\theta$ variable with the Fourier expansion
\begin{equation}
f(\theta,\omega,q,t)=\frac{p(\omega,q)}{2\pi}
\sum_{l=-\infty}^\infty  f_l(\omega,q,t) {\rm e}^{{\rm i}l\theta} 
\label{fourier}
\end{equation}
where $f_l=f_{-l}^*$, $f_{0}=1$.
Inserting this Fourier series into the continuity equation
(\ref{continuity}), an infinite set of
integro-differential equations for the Fourier modes is obtained:
\begin{equation}
\partial_t f_l =-{\rm i} l (\omega+ K q)   f_l  \label{fourier_set}
+\frac{K l}{2} \left[ r^* (1+{\rm i}q) f_{l-1}  -  r (1-{\rm i}q) f_{l+1} \right]
\end{equation}
Note that the order parameter (\ref{z}) is only determined by the first Fourier mode:
\begin{equation}
r^*(t)=\iint_{-\infty}^{\infty} p(\omega,q) f_1(\omega,q,t) ~ {\rm d}\omega ~ {\rm d}q .
\label{z1}
\end{equation}

\section{Ott-Antonsen ansatz}
Recently Ott and Antonsen (OA) found 
an 
ansatz~\cite{OA08},
which is generically \cite{OA09,OHA11} satisfied by the
asymptotic dynamics of the Kuramoto model ($q=0$)---and,
remarkably, of many variations of it, 
see {\it e.g.}~\cite{LOA09,MBS+09,PM09,MP11,LCT10}.
In our case this ansatz takes the form
\begin{equation}
f_l(\omega,q,t)=\alpha(\omega,q,t)^l
\label{ansatz}
\end{equation}
This defines  a family of solutions of eq.~(\ref{fourier_set}) 
with the constraint that $\alpha$ satisfies
\begin{equation}\label{alpha} 
\partial_t \alpha  = -{\rm i}(\omega+ K q) \alpha + \frac{K}{2} \left[  r^*  (1+{\rm i}q)    
 - r (1-{\rm i}q) \alpha^{2}\right] .
\end{equation}
Our simulations indicate that the asymptotic solutions
of the system indeed belong to the OA manifold.

We consider first a family of joint PDFs 
$p(\omega,q)=h(q)g_c(\omega|q)$, with Lorentzian (Cauchy) marginal distribution $h$:
\begin{equation}\label{h}
h(q)=\frac{\gamma/\pi}{(q-q_0)^2+\gamma^2}, 
\end{equation}
and Lorentzian conditional distribution $g_c$:
\begin{equation}\label{pc}
g_c(\omega|q)= \frac{\delta/\pi}{\left[\omega-\omega_0-m(q-q_0)\right]^2+\delta^2} .
\end{equation}
The specific family of PDFs defined by eqs.~(\ref{h}) and (\ref{pc}) allows us to
obtain simple low-dimensional ordinary differential equations for the order parameter dynamics,
and to tune the statistical dependence between $\omega$ and $q$ with the parameter $m$.
(The case $m=0$---$\omega$ and $q$ independent random variables---
was already addressed in \cite{MP11}.)
In the limiting case $\delta\to0$, $g_c$ becomes a Dirac's delta, and this results
in a (deterministic) linear relationship: $\omega=\omega_0+m(q-q_0)$.
The terms $r$ and $r^*$ in eq.~(\ref{alpha}) can be evaluated by means of the residue's theorem
inserting the PDFs (\ref{h}) and (\ref{pc}) in eq.~(\ref{z1}),
and closing the integration paths
in the complex plane.
Concerning variable $\omega$, the integration must be done in
the lower half complex $\omega$-plane because $\alpha$ can be
analytically continued in that region, as occurs in the Kuramoto model, see
\cite{OA08} for details.
In partial fractions
$g_c(\omega|q)=(2\pi {\rm i})^{-1}\{[\omega-\omega_0-m(q-q_0)-{\rm i}\delta]^{-1}
-[\omega-\omega_0-m(q-q_0)+{\rm i}\delta]^{-1}\}$,
and the integration over $\omega$ in eq.~(\ref{z1})  
involves only the value of $\alpha$ at the pole 
$\omega^p=\omega_0-m(q-q_0)-{\rm i}\delta$, see \cite{OA08}.
Hence, eq.~(\ref{z1}) becomes in this particular instance
\begin{equation}
r^*(t)=\int_{-\infty}^{\infty} h(q) ~\alpha(\omega=\omega^p,q,t) ~ {\rm d}q .
\label{z2}
\end{equation}

To evaluate this integral over $q$ we must proceed 
more
cautiously 
to warrant that 
$\alpha$ can be analytically extended into the suitable
half $q$-plane ($q=q_r+{\rm i}q_i$, with either $q_i\ge0$ or $q_i\le 0$).
Equation (\ref{alpha}) for $\alpha$ at $(\omega^p,q)$ is:
\begin{eqnarray}\label{alphat} 
\partial_t \alpha  = &-&{\rm i}[\omega_0-m q_0+ q(m+K)-{\rm i}\delta] \alpha \nonumber \\
&+& \frac{K}{2} \left[  r^*  (1+{\rm i}q)    
 - r (1-{\rm i}q) \alpha^{2}\right] 
\end{eqnarray}
If $\alpha=|\alpha|{\rm e}^{-{\rm i}\psi}$ is analytic, it satisfies the Cauchy-Riemann conditions, 
and this  
can be demonstrated to imply $\partial_{q_r} |\alpha|+\partial_{q_i}|\alpha|\ge 0$.
In consequence, the maximum of $|\alpha|$ is necessarily located on the boundary
(namely, on the integration contour). Under the assumption that $\alpha$ is analytic at $t=0$,
analyticity will hold for all $t>0$ if $\alpha$ remains finite 
because $\alpha$ is the solution of 
the ordinary differential equation (\ref{alpha}) 
(see Theorem~8.4 in Chapt.~1 of \cite{CL}).
Moreover we require $|\alpha|\le 1$ everywhere in the selected half complex $q$-plane;
otherwise the Fourier modes diverge, see eq.~(\ref{ansatz}).
After some algebra, one finds that on the real $q$-axis, eq.~(\ref{alphat}) yields:
\begin{equation}
\partial_t |\alpha| = -\delta |\alpha| +\frac{K}{2} \,
\mbox{Re}\left[r^* {\rm e}^{{\rm i}\psi} (1+{\rm i} q)\right]\left( 1 -|\alpha|^2 \right),
\label{abs_alpha}
\end{equation}
which gives $\partial_t |\alpha| = -\delta < 0$ at $|\alpha|=1$. This implies that
if $|\alpha|\le 1$ at $t=0$, this will hold for all $t<0$. 
On the contour closing at infinity $q=|q|{\rm e}^{{\rm i}\vartheta}$ with  $|q|\to \infty$,
the dominant contributions (of order $|q|$) at $|\alpha|=1$ give:
\begin{equation}
 \partial_t |\alpha|= \left[m+K(1-R \cos\chi)\right] |q| \sin\vartheta,
\label{maestra}
\end{equation}
where $\chi=\psi(q,t)-\Psi(t)$.
If $\partial_t |\alpha|< 0$ is fulfilled in either $\vartheta\in(0,\pi)$
or $\vartheta\in(-\pi,0)$, we can safely choose that path
for the contour closing in the integration of (\ref{z2}).
The problem now is that if $m\ne0$ there are values of $K$ in
eq.~(\ref{maestra}) where the desired relation $\partial_t |\alpha|< 0$
cannot be fulfilled due to the ``uncontrolled'' angle $\chi$.
Instead of ignoring those parameter values 
we shall make the assumption that solutions in the range $R < R_\times$ where
$\partial_t|\alpha|<0$ at $|\alpha|=1$ is fulfilled (choosing the appropriate half-plane),
can be correctly studied within this framework. Thus,
eq.~(\ref{maestra}) dictates that the analysable range of $R$ is
bounded by 
\begin{equation}\label{rx}
 R_\times=\min(1,|1+m/K|) ,
\end{equation}
what in particular implies that, in principle,
the stability of incoherence ($R=0$) can be always determined, 
save at $K=-m$ ($R_\times=0$).
We must take 
$\vartheta\in(-\pi,0)$ if $m+K>0$, and
$\vartheta\in(0,\pi)$ if $m+K<0$, for the
closing of the integration contour in eq.~(\ref{z2}).
Thus, the order parameter is determined by the value of $\alpha$ at the poles
\begin{equation}
r^*(t)=\alpha(\omega=\omega^p,q=q^p,t) ,
\label{z3}
\end{equation}
with $q^p=q_0-{\rm i}\gamma$ for $m+K>0$,
and $q^p=q_0+{\rm i}\gamma$ for $m+K<0$. Equation (\ref{z3}) yields the relations $R(t)=|\alpha(\omega^p,q^p,t)|$
and $\Psi(t)=\psi(\omega^p,q^p,t)$, and hence it suffices to study eq.~(\ref{alpha}) at
$(\omega,q)=(\omega^p,q^p)$.

Recalling that $q^p=q_0\mp {\rm i} \gamma$, and $\omega^p=\omega^0+m(q^p-q_0)-{\rm i}\delta$, we obtain
that the modulus and the phase of the order parameter (inside the OA manifold)
obey Stuart-Landau equations:
\begin{eqnarray}\label{rdot}
 \dot R &=& \left[-\delta\mp m \gamma + \frac{K}{2}(1\mp\gamma)(1-R^2) \right]R\\
\dot \Psi &=& \omega_0 + \frac{K}{2}q_0(1-R^2) \label{psidot}
\end{eqnarray}
Remarkably, the radial dynamics does not depend on $q_0$, something that
stems from the peculiarities (pointed out in \cite{MP11}) of the Lorentzian distribution.

In the incoming paragraphs we present separately the cases of positive and negative $m$,
as these two cases yield qualitatively different results.

\section{Positive dependence ($m>0$)}

In this case, eq.~(\ref{rx}) implies $R_\times=1$ 
for $K/m\ge -\tfrac{1}{2}$, and $R_\times<1$ for $K/m< -\tfrac{1}{2}$. In the latter region we
cannot solve the problem completely within the OA framework because possible attractors
with $R\in [R_\times,1]$ are not captured by the theory.

\begin{figure}
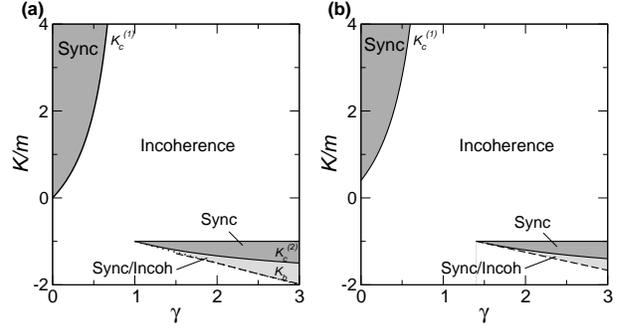

\onefigure[width=80mm]{fig1.eps}
\caption{Phase diagrams for positive dependence between $\omega$ and $q$:
eq.~(\ref{pc}) with $m>0$. 
(a) Purely linear dependence $\delta=0$. 
(b) $\delta=0.2 m$.
The solid lines correspond to the loci of bifurcations
given by the analytic formulas (\ref{kc1+}), (\ref{kc2+}), and $K=m$ (see text).
The dotted line is a bound of the region of bistability given by eq.~(\ref{kb}).
Dashed lines are obtained from numerical simulations with  $N=$ $2000$ (a), $40000$ 
(b) oscillators. Our numerics showed good agreement with 
boundaries (\ref{kc1+}) and (\ref{kc2+}), data not shown.
In the simulations we took $\{\omega_j,q_j\}_{j=1,\ldots,N}$ deterministically to represent
$p(\omega,q)$ given by eqs.~(\ref{h}) and (\ref{pc}); the selected
parameters were $m=1$ and $\omega_0=q_0=\tfrac{1}{2}$.}
\label{fig.1}
\end{figure}

If $K>-m$, 
the signs ``$\mp$'' in eq.~(\ref{rdot}) must be replaced by ``$-$''.
It can be easily seen that incoherence is stable everywhere, except above the line:
\begin{equation}\label{kc1+}
K_c^{(1)}=\frac{2(m\gamma+\delta)}{1-\gamma}  \qquad \mbox{with $\gamma<1$}
\end{equation}
A phase diagram for two values of $\delta$ can be seen in fig.~\ref{fig.1}.
At $K_c^{(1)}$ a supercritical bifurcation gives rise to a partially synchronized solution with
\begin{equation}\label{r2k}
R^2=\frac{K-K_c}{K} .
\end{equation}
Remarkably, this formula coincides with the one obtained in
the standard Kuramoto model \cite{Kur84} (recovered at $\gamma=q_0=0$).

If $K<-m$, one must replace ``$\mp$'' by ``$+$'' in eq.~(\ref{rdot}). The resulting
equation predicts the incoherence to be unstable 
in the wedge-shaped region between $K=-m$  and
\begin{equation}\label{kc2+}
K_c^{(2)}= - \frac{2(m\gamma-\delta)}{1+\gamma} \qquad \mbox{with $\gamma >1+\frac{2\delta}{m}$}. 
\end{equation}
If $\omega$ and $q$ are let to be progressively less statistically dependent
($m \to 0$), the tip of this region goes to $\gamma=\infty$. Thus,
the interval of $\gamma$ where incoherence is stable for all $K$ becomes infinite
as $m\to 0$, in consistence with our result in \cite{MP11} for the independent case ($m=0$).
Inside the wedge-like region where incoherence is unstable we can presume
---and confirm numerically---
the existence of a stable partially synchronized solution (with $R\ge R_\times$). Moreover, as
the instability of incoherence at $K_c^{(2)}$ is subcritical,
we can infer the existence of a region of bistability incoherence-synchronization
below this line. The unstable solution with $R>0$ appearing at $K_c^{(2)}$ can
be analytically determined up to 
\begin{equation}\label{kb}
 K_b=-\frac{m^2(1+\gamma)}{2(m+\delta)}
\end{equation}
where it acquires an $R$ larger than $R_\times$. 
Hence $K_b$ is a bound (surprisingly tight) for the region of bistability, see fig.~1.

\section{Negative dependence ($m<0$)}
In the case of negative $m$, eq.~(\ref{rx}) tells us that our eqs.~(\ref{rdot}) and
(\ref{psidot}) apply to all $R$ values when $K/|m|\le \tfrac{1}{2}$,
while otherwise their validity only holds in a certain range $R < R_\times$.
In contrast to the case of positive $m$, the phase diagram undergoes
several transformations as the ratio between $\delta$ and $|m|$ varies.
Next we describe the three main situations separately, see fig.~2.

\begin{figure}
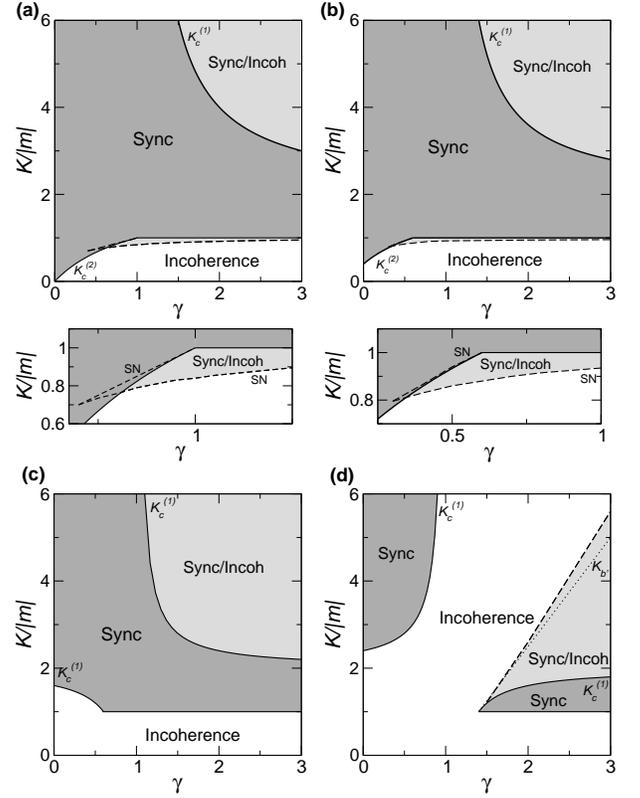

\onefigure[width=80mm]{fig2a.eps}
\onefigure[width=80mm]{fig2b.eps}
\caption{Phase diagram for $m<0$ with $\delta=0$  (a), $0.2|m|$ (b), $0.8|m|$ (c), and $1.2|m|$ (d).
Solid lines are analytical predictions (tested by numerical simulations). The
dashed lines are obtained directly from numerical simulations with $m=-1$, $q_0=\tfrac{1}{2}$ and $N=$ $2000$.
The dotted line in panel (d) is a bound of the region of bistability 
given by eq.~(\ref{kbp}).Small panels show magnified regions of the the phase diagrams (a)
and (b).}
\label{fig.2}
\end{figure}

\subsection{Case I: $0\le\delta<|m|/2$; fig.~2(a,b)} If $K>|m|$ incoherence
is stable only above the line
\begin{equation}\label{kc1-}
K_c^{(1)}=\frac{2(-|m|\gamma+\delta)}{1-\gamma}  ,\qquad \mbox{with $\gamma>1$} 
\end{equation}
where an unstable solution branches off incoherence obeying relation (\ref{r2k}).
As presumable, a region of bistability between incoherence and synchronization (with $R\ge R_\times$) is found.
For $K<|m|$ incoherence is stable everywhere except above the line
\begin{equation}\label{kc2-}
K_c^{(2)}=\frac{2(|m|\gamma+\delta)}{1+\gamma}  ,\qquad \mbox{with $\gamma< 1-\frac{2\delta}{|m|}$} 
\end{equation}
where it undergoes a supercritical bifurcation.

Our numerical simulations reveal that a stable coherent solution
exists below $K/|m|=1$ in the region of stable incoherence, see
bottom panels of fig.~\ref{fig.2}(a,b).
For $\delta=0$, this solution is continuation of a fully
synchronized solution existing at $K/|m|=1$ with
$R=\int_{-\infty}^\infty h(q)(1+q^2)^{-1/2} {\rm d}q$.
This solution depends on $|q_0|$, and in consequence the region of
bistability is also $|q_0|$-dependent.
The bifurcation scenario is consistent with
two saddle-node (SN) bifurcations emanating
from a (codimension-2) cusp point.

\subsection{Case II: $|m|/2<\delta<|m|$; fig.~2(c)}
The only relevant bifurcating lines (in addition to $K=|m|$) are given by
$K_c^{(1)}$ in
eq.~(\ref{kc1-}) with a left branch emanating from the $K$-axis and existing
up to $\gamma=2\delta/|m|-1$, and a right branch existing above $\gamma=1$.
At the left branch of $K_c^{(1)}$ the bifurcation from incoherence is supercritical,
while it is subcritical at the right branch.

\subsection{Case III: $\delta>|m|$; fig.~2(d)}
At $\delta=|m|$ the locus of $K_c^{(1)}$ [eq.~(\ref{kc1-})] reorganizes giving rise to the phase diagram
for $\delta>|m|$ shown in fig.~2(d). A wedge-like region of unstable incoherence 
between $K_c^{(1)}$ and $K=|m|$ exist above $\gamma=2\delta/|m|-1$. This means that in the
limit $m\to0^-$ this region disappears and the phase diagram becomes the one found in \cite{MP11}
in the independent case ($m=0$), as expected.
The fact that the right branch of $K_c^{(1)}$ corresponds to a subcritical bifurcation
results in a region of bistability. This region cannot be analytically determined,
though a lower bound for its upper border can be calculated finding 
where an unstable solution (with $0<R<R_\times$) exists. We obtain the line
\begin{equation}
 K_{b'}=\frac{m^2 (1-\gamma)}{2(|m|-\delta)} \qquad \mbox{with $\gamma> \frac{2\delta}{|m|}-1$}
\label{kbp}
\end{equation}
shown as a dotted line in fig.~\ref{fig.2}(d), which is
a good estimation of the upper border of the bistable region.
 
Remarkably, the phase diagrams for negative dependence 
differ significantly from those obtained
for positive dependence (fig.~\ref{fig.1}). With negative $m$,
synchrony becomes more dominant in the
phase diagram, in consonance with the numerical observations
made in \cite{BMK03}.

\section{Linear stability analysis}
With general distributions the residue's theorem cannot be used.
We can nevertheless follow Strogatz and Mirollo \cite{SM91} and calculate 
the linear stability threshold of the incoherent state.
This allows to know how much our
results for the Lorentzian $h(q)$ are applicable to other distributions,
what is always a concern when applying the OA theory \cite{LCT10,MP11}. 
Our analysis is not completely rigorous but permits to understand the
results of the numerical simulations.

In the incoherent state, all Fourier modes (save $f_0$)
vanish: $f_{l\ne0}=0$. Equations (\ref{fourier_set}) for the
Fourier modes indicate that at the lowest order
only the first Fourier mode $l=\pm1$ is relevant, and it obeys:
\begin{eqnarray} \label{continuity2}
\frac{\partial f_1}{\partial t}=&-&{\rm i} (\omega+q K) f_1 \\  
&+& \frac{K}{2}(1+{\rm i} q) \iint_{-\infty}^\infty f_1(\omega',q',t) p(\omega',q') \, {\rm d}\omega'\, {\rm d}q' \nonumber
\end{eqnarray}
The linear operator in the right hand side has a linear spectrum of
eigenvalues $\lambda$. If $f_1(\omega,q,t)=b(\omega,q)\exp(\lambda t)$
is inserted into eq.~(\ref{continuity2}) and the trivial solution $b=0$ is discarded,
we get:

\begin{equation}
\frac{2}{K}=
\iint_{-\infty}^\infty \frac{1+ {\rm i} q }{\lambda + {\rm i}(\omega+q K)} p(\omega,q) \, {\rm d}\omega\, {\rm d}q
\label{original}
\end{equation}
Defining $\lambda=\lambda_r+ {\rm i}\lambda_i$, one finds that the imaginary part of 
eq.~\eqref{original} has always a solution $\lambda_i=-\omega_0$ at the
stability threshold ($\lambda_r\to0^+$)
if the distribution $h(q)$ is centred at zero.  
As an important example, let us mention the case of
Gaussian PDFs:
$$
h(q)=\frac{1}{\sqrt{2\pi}\nu} {\rm e}^{-\frac{q^2}{2\nu^2}}, 
\qquad g_c(\omega|q)= \frac{1}{\sqrt{2\pi}\sigma} {\rm e}^{-\frac{(\omega-\omega_0-m q)^2}{2\sigma^2}} .
$$
(hereafter we take $\omega_0=0$ as this can always be achieved going into a
rotating framework).
We obtain an equation for the stationary ($\lambda_i=0$)
instability of incoherence:
\begin{equation}
\frac{2}{K_c^s}=\sqrt{\frac{\pi}{2\nu^2(K_c^s+m)^2+2\sigma^2}} + \frac{\nu^2 (K_c^s+m)}{\nu^2(K_c^s+m)^2+\sigma^2} 
\label{kcs}
\end{equation}
For $m=0$ a simple analytic solution for $K_c^s$ can be found \cite{MP11}; otherwise
$K_c^s$ is the solution of a fourth-order polynomial.
In the next section we show that sometimes
complex eigenvalues ($\lambda_i\ne0$) may also destabilize incoherence,
and hence using eq.~(\ref{kcs}) we take the risk of missing nonstationary instabilities.

\section{Linear dependence between $\omega$ and $q$}
If there exists a purely linear dependence of $\omega$ on $q$, $\omega_j=m(q_j-q_0)$,
general expressions for the stability threshold of incoherence can be obtained
if $q_0=0$.
With this latter choice the system possesses
reflection symmetry $(\theta_j,\omega_j,q_j)\to(-\theta_j,-\omega_j,-q_j)$, in addition to
the rotational symmetry $\theta_j\to\theta_j+\phi$.
Inserting the pdf $p(\omega,q)=h(q)\delta(\omega-m q)$
into eq.~(\ref{original}), 
and taking the limit $\lambda_r\to0^+$, we obtain:
\begin{equation}
\label{interm}
 \frac{2}{K_c^s}=\frac{\pi h(0)}{|K_c^s+m|}+\frac{1}{K_c^s+m} .
\end{equation}
Solving this equation for $K_c^s$ gives the boundaries:
\begin{equation}\label{kc}
K_c^s= 
\begin{cases} \frac{2m}{\pi h(0) -1} & \text{if $K_c^s>-m$,}
\\
\frac{-2m}{\pi h(0) +1} &\text{if $K_c^s< -m$.}
\end{cases}
\end{equation}
The linear stability analysis 
permits to determine at which
side of the bifurcation the incoherent state is unstable. This is an 
indirect indication that 
there must exist a horizontal bifurcation line at $K=-m$,
exactly like in figs.~\ref{fig.1}(a) and \ref{fig.2}(a)
for Lorentzian $h(q)$. Moreover eq.~(\ref{kc}) agrees with the
analytical and numerical results obtained in figs.~\ref{fig.1}(a)
and \ref{fig.2}(a) for positive and negative $m$, respectively.
Note also that reflection symmetry makes the stationary
instability at $K_c^s$ to be a (circle-)pitchfork bifurcation.
We also report next the results obtained with Gaussian $h(q)$:

\begin{figure}
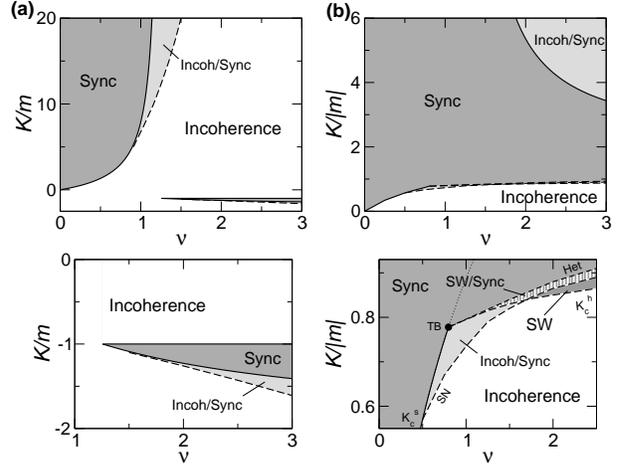

\onefigure[width=80mm]{fig3.eps}
\caption{Phase diagrams of model \eqref{model} with Gaussian $h(q)$ and $g_c=\delta(\omega-mq)$
for  $m>0$ (a)  and $m<0$ (b). 
Solid lines (numerically tested) correspond to eq.~(\ref{kc}) [and $K=-m$ in (a)].
Dashed lines are obtained from numerical simulations with $|m|=1$ and $N=4000$.
Small panels show magnified regions of the the phase diagrams (a)
and (b).}
\label{fig.gaussian}
\end{figure}

\subsection{Positive $m$} The result of our numerical simulations with Gaussian $h(q)$
and $m>0$ is presented in fig.~3(a), and confirms the soundness of eq.~(\ref{kc}).
In contrast to the case of Lorentzian $h(q)$ in fig.~\ref{fig.1}(a),
a region of bistability between synchronization and incoherence
exists at large $K$. This can be understood taking the limit $K\to\infty$,
in eq.~(\ref{model}) and performing a self-consistence analysis
\`a la Kuramoto, see \cite{MP11}. A solution branches off from incoherence
at $\pi h(0)=1$ increasing the value of $\nu$, a scenario of subcritical
bifurcation coherent with the observed bistability.
The orientation of this branch is intrinsic to the form of $h(q)$
and is independent of the value of $m$.
For 
distributions with a sharp peak, like the triangular or Laplace distributions,
the bifurcation is supercritical [and the phase diagram will be slightly
different from that in fig.~\ref{fig.gaussian}(a)].
The Lorentzian distribution is marginal and finite-$K$ effects 
make the bifurcation to be supercritical 
for $m\ge 0$ and subcritical for $m<0$.

\subsection{Negative $m$} The numerical results for Gaussian $h(q)$,
shown in fig.~\ref{fig.gaussian}(b),
 indicate that eq.~(\ref{kc}) predicts everywhere
the correct boundaries for stable incoherence,
except in a region close to $K=|m|$ (see bottom panel).
There incoherence undergoes a Hopf bifurcation at $K_c^h$,
a bifurcation line that emanates from 
a double zero eigenvalue (Takens-Bogdanov) point located on 
$K_c^s$ at $\nu_{TB}=\sqrt{2/\pi}$. (This stems from a degeneracy at
$\pi h(0)=-\int h'(q) q^{-1} {\rm d}q$.) It is remarkable that the Hopf bifurcation
gives rise to a standing wave (SW) consisting of two counter-rotating clusters
of locked oscillators. In the standard
Kuramoto model the SW cannot arise in unimodal distributions of $\omega$,
but it is typical of bimodal distributions with well separated peaks \cite{ABP+05,MBS+09,PM09}.
The other lines in the bottom panel of fig.~\ref{fig.gaussian}(b)
are (twin) saddle-node bifurcations (SN) emanating
from a degenerate-pitchfork point, and a heteroclinic connection (Het)
born at TB.

Taking $q_0\ne0$
breaks the reflection symmetry and the phase diagram should exhibit structures already
found in the Kuramoto model with bimodal non-symmetric distribution \cite{acebron98}
or unbalanced interacting subpopulations \cite{MKB04}.

\section{Conclusions} 
Our work is a natural step in the development,
initiated by Winfree and Kuramoto,
of realistic solvable phase models, as simplifications of the mean-field
complex Ginzburg-Landau equation
\cite{Kur75,SK86} or in other set-ups \cite{Win67,boni98,AS01}.
The model analysed in this Letter widens the scope of the Kuramoto model
by admiting shear diversity.
Shear is a generic feature of oscillators with
particular relevance in ensembles of limit-cycles close to 
collision with a saddle point
(saddle-loop bifurcation)  \cite{han}. These systems may be good  candidates to 
observe the phenomena reported here.

Considering a broad but still 
reasonably simple family of joint distributions $p(\omega,q)$, we 
have found that the sign and magnitude of $m$, 
controlling the dependence between
the natural frequencies and the shears, has a profound impact
on the phase diagrams. Synchronization
is prevalent for negative $m$, whereas incoherence prevails
if $m$ is positive (or zero \cite{MP11}). A certainly interesting 
line of future work would be to investigate the effect   
of other dependences between $\omega$ and $q$ on the synchronization
phase diagrams.

Finally, this work can also give hints about the validity of the OA ansatz in systems
with distributed parameters~\cite{PR11}. Why
distributing $q$ is so amenable to analysis?

\acknowledgments
Financial support from the MICINN (Spain) under
project No.~FIS2009-12964-C05-05 is acknowledged.


\begin{thebibliography}{10}
\expandafter\ifx\csname url\endcsname\relax\def\url#1{\texttt{#1}}\fi

\bibitem{Win67}
\Name{Winfree A.~T.} \REVIEW{J. Theor. Biol.}{16}{1967}{15}.

\bibitem{Kur84}
\Name{Kuramoto Y.} \Book{Chemical Oscillations, Waves, and Turbulence}
  ({S}pringer-{V}erlag, Berlin) 1984.

\bibitem{PRK01}
\Name{Pikovsky A.~S., Rosenblum M.~G. \and Kurths J.} \Book{Synchronization, a
  Universal Concept in Nonlinear Sciences} (Cambridge University Press,
  Cambridge) 2001.

 \bibitem{MMZ04}
 \Name{Manrubia S.~C., Mikhailov S.~S. \and Zanette D.~H.} \Book{Emergence of
   Dynamical Order} (World Scientific, Singapore) 2004.

\bibitem{ABP+05}
\Name{Acebr\'on J.~A. \etal} \REVIEW{Rev. Mod. Phys.}{77}{2005}{137}.

\bibitem{Kur75}
\Name{Kuramoto Y.} \Book{Self-entrainment of a population of coupled non-linear
  oscillators} in \Book{International Symposium on Mathematical Problems in
  Theoretical Physics}, edited by \Name{Araki H.} Vol.~39 of \emph{Lecture
  Notes in Physics} (Springer, Berlin) 1975 pp. 420--422.


\bibitem{pere} \Name{Wiesenfeld K., Colet P. \and Strogatz S.~H.} \REVIEW{Phys. Rev. Lett.}{76}{1996}{404}.

\bibitem{uchida} \Name{Uchida N. \and Golestanian R.} \REVIEW{EPL}{89}{2010}{50011}.

\bibitem{mertens} \Name{Mertens D. \and Weaver R.} \REVIEW{Phys. Rev. E}{83}{2011}{046221}.

\bibitem{cross} \Name{Cross M.~C. \etal} \REVIEW{Phys. Rev. Lett.}{93}{2004}{224101};
\REVIEW{Phys. Rev. E}{73}{2006}{036205}.

\bibitem{aizawa76}
\Name{Aizawa Y.} \REVIEW{Prog. Theor. Phys.}{56}{1976}{703}.
\Name{Shiino M. \and Frankowicz M.} \REVIEW{Phys. Lett. A}{136}{1989}{103}.
\Name{{Matthews} P.~C. \and {Strogatz} S.~H.} \REVIEW{Phys. Rev. Lett.}{65}{1990}{1701}.
\Name{{Matthews} P.~C., Mirollo R.~E. \and {Strogatz} S.~H.} \REVIEW{Physica D}{52}{1991}{293}.
\Name{{de Monte} S. \and d'Ovidio F.} \REVIEW{Europhys. Lett.}{58}{2002}{21}.

\bibitem{SK86}
\Name{Sakaguchi H. \and Kuramoto Y.} \REVIEW{Prog.~Theor.~Phys.}{76}{1986}{576}.

\bibitem{MP11}
\Name{Montbri\'o E. \and Paz\'o D.} \REVIEW{Phys. Rev. Lett.}{106}{2011}{254101}.

\bibitem{BMK03}
\Name{Blasius B., Montbri\'o E. \and Kurths J.} \REVIEW{Phys. Rev. E}{67}{2003}{035204}.
\Name{Montbri\'o E. \and Blasius B.} \REVIEW{Chaos}{13}{2003}{291}.

\bibitem{OA08}
\Name{Ott E. \and Antonsen T.~M.} \REVIEW{Chaos}{18}{2008}{037113}.

\bibitem{OA09}
\Name{Ott E. \and Antonsen T.~M.} \REVIEW{Chaos}{19}{2009}{023117}.

\bibitem{OHA11}
\Name{Ott E., Hunt B.~R. \and Antonsen T.~M.} \REVIEW{Chaos}{21}{2011}{025112}.

\bibitem{LOA09}
\Name{Lee W.~S., Ott E. \and Antonsen T.~M.} \REVIEW{Phys. Rev. Lett.}{103}{2009}{044101}.
\Name{Hong H. \and Strogatz S.~H.} \REVIEW{Phys. Rev. Lett.}{106}{2011}{054102}.

\bibitem{MBS+09}
\Name{Martens E.~A. \etal} \REVIEW{Phys. Rev. E}{79}{2009}{026204}.

 \bibitem{PM09}
 \Name{Paz\'o D. \and Montbri\'o E.} \REVIEW{Phys. Rev. E}{80}{2009}{046215}.

\bibitem{LCT10}
\Name{Lafuerza L.~F., Colet P. \and Toral R.} \REVIEW{Phys. Rev. Lett.}{105}{2010}{084101}.

\bibitem{CL}
\Name{Coddington E.~A. \and Levinson N.} \Book{Theory of Ordinary Differential
  Equations} (McGraw-Hill, New York) 1955.

\bibitem{SM91}
\Name{Strogatz S.~H. \and Mirollo R.~E.} \REVIEW{J. Stat. Phys.}{63}{1991}{613}.

\bibitem{acebron98}
\Name{Acebr\'on J.~A. \etal} \REVIEW{Phys.
  Rev. E }{57}{1998}{5287}.


\bibitem{MKB04}
\Name{Montbri\'o E., Kurths J. \and Blasius B.} \REVIEW{Phys. Rev. E}{70}{2004}{056125}.

\bibitem{boni98}
\Name{Bonilla L.~L. \etal} \REVIEW{Phys. Rev. Lett.}{81}{1998}{3643}.


\bibitem{AS01}
\Name{Ariaratnam J.~T. \and Strogatz S.~H.} \REVIEW{Phys. Rev. Lett.}{86}{2001}{4278}.

\bibitem{han}
\Name{Han S.~K., Kurrer C. \and Kuramoto Y.} \REVIEW{Phys. Rev. Lett.}{75}{1995}{3190}.

\bibitem{PR11}
\Name{Pikovsky A. \and Rosenblum M.} \REVIEW{Physica D}{240}{2011}{872 }.

\end{thebibliography}

\bibliographystyle{eplbib}

\end{document}